\documentstyle[12pt]{article}
\normalsize

\def\a{\alpha}
\def\b{\beta}

\def\g{\gamma}

\def\s{\sigma}

\def\be{\begin{equation}}
\def\ee{\end{equation}}

\def\Hat#1{\rlap{\kern.10em$\widehat{\phantom G}$}#1}
\def\HAt#1{\rlap{\kern.05em$\widehat{\phantom G}$}#1}

\def\cap#1{\rlap{\kern.1em$\widehat{\phantom{G\vrule height.8em}}$}#1{}}
\def\Cap#1{\rlap{\kern.05em$\widehat{\phantom{G\vrule height.8em}}$}#1{}}

\let\oldtheequation=\theequation
\def\doteqs#1{\setcounter{equation}{0}
            \def\theequation{{#1}.\oldtheequation}}
\newcounter{sxn}
\def\sx#1{\addtocounter{sxn}{1} \bigskip\medskip \goodbreak
\noindent{\large\bf
\centerline{\thesxn.~~#1}} \nobreak \medskip}
\def\sxn#1{\sx{#1} \doteqs{\thesxn}}
\newcounter{axn}
\def\ax#1{\addtocounter{axn}{1} \bigskip\medskip \goodbreak
\noindent{\large\bf
{\Alph{axn}.~~#1}} \nobreak \medskip}
\def\axn#1{\ax{#1} \doteqs{\Alph{axn}}}
\newcommand{\ba}{\begin{eqnarray}}
\newcommand{\ea}{\end{eqnarray}}
\def\br{\begin{thebibliography}{abc}}
\def\er{\end{thebibliography}}

\date{}
\begin{document}
\bibliographystyle{unsrt}
\footskip 1.0cm
\thispagestyle{empty}
\setcounter{page}{0}
\begin{flushright}
Napoli DSF-T-28/93,INFN-NA-IV-28/93\\
Syracuse SU-4240-548\\
September 1993
\end{flushright}
\begin{center}{DISCRETIZED LAPLACIANS ON AN INTERVAL \\
AND THEIR RENORMALIZATION GROUP\\ }
\vspace*{6mm}
{\large G. Bimonte $^{(1,2)}$, E .Ercolessi $^{(1)}$ \\
and \\
P. Teotonio-Sobrinho $^{(1)}$}

\newcommand{\bc}{\begin{center}}
\newcommand{\ec}{\end{center}}
\vspace*{5mm}
 1){\it Department of Physics, Syracuse University,\\
Syracuse, NY 13244-1130, USA}.\\
\vspace*{4mm}
 2){\it Dipartimento di Scienze Fisiche dell' Universit\`a di Napoli,\\
    Mostra d'Oltremare pad. 19, 80125 Napoli, Italy,\\
    and\\
 Istituto Nazionale di Fisica Nucleare, Sezione di Napoli,\\
    Mostra d'Oltremare pad. 19, 80125 Napoli, Italy}.
\ec

\vspace*{5mm}

\normalsize
\centerline{\bf ABSTRACT}

\vskip.3cm

The Laplace operator admits infinite self-adjoint extensions when considered
on a segment of the real line. They have different domains of essential
self-adjointness characterized by a suitable set of boundary conditions on
the wave functions. In this paper we show how to recover these extensions by
studying the continuum limit of certain discretized versions of the Laplace
operator on a lattice. Associated to this limiting procedure,
there is a renormalization flow in the finite dimensional parameter space
describing the dicretized operators.
This flow is shown to have infinite fixed points, corresponding to the
self-adjoint extensions characterized by scale invariant boundary conditions.
The other extensions are recovered by looking at the other trajectories of
the flow.
\newpage
\setcounter{page}{1}

\sxn{Introduction}

\indent

Since the time of their introduction by Wilson \cite{wil}, lattice field
theories have proven to be a successfull tool in the study of quantum field
theories. In this approach, one replaces the space-time continuum (or rather,
in the Hamiltonian version of the method \cite{kog},
the space continuum ) with a discrete lattice. Correspondingly, the action (or
the Hamiltonian) describing the continuum theory is replaced by a
discretized one. The spacing $a$ of the lattice then provides a natural
cut-off for the theory and the continuum limit is recovered by taking $a$ to
zero. This process is governed by the renormalization group, which
dictates how the parameters on which the discretized action depends have to
evolve with $a$ so that, for $a$ sufficiently small, the physics becomes
insensitive to the presence of the lattice spacing; when this happens,
one says that the continuum limit exists. Unfortunatly, for most of the
field theories of interest, the analysis of the renormalization group is too
complicated to be carried out exactly and can be attacked only within
perturbation theory.

In this paper we discuss a non trivial illustration of the ideas of Wilson
already in the context of quantum mechanics. The idea of renormalization has
been applied already to quantum mechanics by other authors \cite{Thor},
\cite{gupta}, but in a different way from that pursued here. In particular, in
\cite{gupta}, it is shown how the use of renormalization allows one to make
physical sense of the inverse square potential, in one space dimension, by
rendering finite the energy of the ground state, which would otherwise be
infinitely negative.

What we consider here is simply a free particle confined to move on a finite
interval. Its Hamiltonian is just a free Laplacian and the only non-trivial
dynamical aspect of the system is the
behaviour of the particle at the end-points. This behaviour is described,
in mathematical terms, by a set of boundary conditions to be
imposed on the wave functions. The most general choice of the boundary
conditions compatible
with the requisite of self-adjointness for the Hamiltonian is parametrized by
a $U(2)$ matrix.

Upon discretisation, the interval is replaced by $N$
points and the Laplacian by the classic finite difference operator already
used by Laplace. In the continuum limit, one recovers the self-adjoint
extension of the Laplacian characterized by periodic boundary conditions for
the wave functions. One can then wonder whether it is also possible to recover
all the other self-adjoint extensions as continuum limits of some suitably
modified finite difference operators. Indeed, after noticing that the
classical finite difference operator is well defined only away from the end
points of the interval, in order to define it everywhere, one is led to
consider its most general symmetric extensions.
The extended operators depend on 4 real parameters, which are the discrete
analogue of the $U(2)$ matrix found earlier in the continuum. The continuum
limit is then obtained by taking $N$ to infinity. The corresponding
four-dimensional renormalization flow turns out to be exactly solvable and
neverthless non trivial. We find and infinite number of fixed points which are
related to the self-adjoint  extensions of the Laplacian characterized by a set
of scale-invariant boundary conditions.
The other non-trivial trajectories, which in the limit approach the
fixed points, correspond instead to the remaining choices of the boundary
conditions.

The paper is organized as follows. Section 2 reviews some basic facts
about the self adjiont extensions of the Laplacian on an interval. In Section 3
we introduce the finite difference operators which represent the discretised
version of the Laplacian with arbitrary boundary conditions for an interval and
in Section 4 we discuss the related eigenvalue problem. The continuum limit is
discussed in Section 5. Finally, the Appendix contains a detailed analysis
of the eigenvalue equations associated to the finite difference operators
introduced in Section 3.

\sxn{Self-Adjoint extensions of the Laplacian on an interval.}

\indent

Let us consider the interval of the real line $[0, 2 \pi]$ and let
${\cal L}^2 ([0, 2\pi])$ be the Hilbert space of square integrable functions
over $[0, 2 \pi]$ with inner product given by:
\be
\langle \phi, \psi \rangle~=~\int_0^{2\pi} dx \phi^* \psi~. \label{1}
\ee
Consider now the Laplacian operator: $-\frac{d^2}{dx^2}$. As it
stands it is only a formal differential operator. In order to obtain a
well defined operator in ${\cal L}^2([0, 2\pi])$ it is necessary to specify
the domain of functions on which it acts. As a starting domain, consider the
set ${\cal D}^0$ of functions which are twice differentiable and vanish
at the end-points together with their first derivatives:
\be
{\cal D}^0~\equiv~\{ \psi(x)~:~\psi(x) \in C^2[0, 2\pi]~;~\psi(0)=\psi(2\pi)=
\partial_x \psi(0)=\partial_x \psi(2 \pi)=0~\} ~. \label{2}
\ee
Let us call $\Delta^0$ the operator obtained by restricting the Laplacian to
${\cal D}^0$. $\Delta^0$ is symmetric, as a simple computation shows:
$$
\langle \psi^0, \Delta^0 \phi^0 \rangle = -\int_0^{2\pi} \psi^{0*}
\partial_x^2 \phi^0 dx=-\psi^{0*} \partial_x \phi^0 |_0^{2\pi} +
\partial_x \psi^{0*}  \phi^0 |_0^{2\pi}
+\langle  \Delta^0 \psi^0,  \phi^0 \rangle =
$$
\be
=\langle  \Delta^0 \psi^0,  \phi^0 \rangle~~~~~ \forall \phi^0, \psi^0 \in
{\cal D}^0. \label{3}
\ee
Consider now the adjoint $\Delta^{0 \dagger}$ of $\Delta^0$. It is defined
by the equation:
\be
\langle  \Delta^{0 \dagger} \psi,  \phi^0 \rangle~=~
\langle \psi, \Delta^0 \phi^0 \rangle ~~~~~~\forall \phi^0 \in {\cal D}^0.
\label{4}
\ee
The domain of $\Delta^{0 \dagger}$ is defined to be the set ${\cal D}
^{0 \dagger}
\subset {\cal L}^2 ([0, 2\pi])$ for which equation (\ref{4}) is fulfilled, for
all $\phi^0 \in {\cal D}^0$. It is easily seen that
\be
{\cal D}^0 \subset {\cal D}^{0 \dagger} ~. \label{5}
\ee
In fact, eq. (\ref{4}) is satisfied by all twice differentiable functions
$\psi$, regardless of their boundary values or of the boundary values of their
derivatives. This shows that $\Delta^0$, though symmetric, is not self-adjoint
(s.a.).
In order to make it s.a. it is necessary to extend its domain from ${\cal D}^0$
to a larger one ${\cal D}$. If it is possible to do it and in how many
different ways, is established using the deficiency index theorem
\cite{reed}. This theorem can be stated as follows. Let $A^0$ be a symmetric
operator on some Hilbert space
${\cal H}$, with domain ${\cal D} (A^0)$ and let $A^{0 \dagger}$ be its
adjoint.
Let now $n_+~,~n_-$ be the number of linearly independent solutions of the
equation
\be
A^{0 \dagger} \psi^{(\pm)}_j = \pm i \psi^{(\pm)}_j~~,~~j=1,\cdots,n_{\pm}~.
\label{6}
\ee
$n_+, n_-$ are called respectively positive and negative
deficiency indices. Then, $A^0$ admits s.a. extensions if and only if
$n_+=n_-=n$ and they can be all obtained in the following way. Let $U \equiv
\{U_{ij}\}$ be an arbitrary $n \times n$ unitary matrix. Consider now the
domain:
\be
{\cal D} _U \equiv {\cal D}^0 \bigoplus {\rm span} \{ \psi^{(+)}_i +
U_{ij} \psi^{(-)}_j \}. \label{7}
\ee
The operators
$$
A_U~:~{\cal D} _U \rightarrow {\cal H}
$$
\be
A \psi= A^0 \psi^0 + i c_j(\psi^{(+)}_j -
U_{jk} \psi^{(-)}_k), \label{8}
\ee
where
$$
\psi = \psi^0 +  c_j(\psi^{(+)}_j +
U_{jk} \psi^{(-)}_k)
$$
represent, for each $U$, a different extension of $A^0$ and it can be shown
that they are all essentially s.a.. Moreover it can be proven that all
possible s.a. extensions of $A^0$ can be obtained in this way, by taking
different unitary matrices $U$ in equations (\ref{7}),(\ref{8}).
\indent

In the case of $\Delta^0$ it is easy to check that $n_+=n_-=2$. $\Delta^0$
then admits an $U(2)$ infinity of s.a. extensions, characterized by domains
of the form (\ref{7}). Now, it can be shown that, for all $U \in U(2)$
there exists a set of homogeneous
linear boundary conditions (b.c.) which uniquely characterizes
${\cal D} _U$:
\be
{\cal D} _U~=~\{ \psi~:~a_{i1}^U\psi^{\prime}(0)+a_{i2}^U \psi^{\prime}(2 \pi)
+a_{i3}^U \psi(0)+a_{i4}^U\psi(2\pi)=0~~~i=1,2\}~. \label{9}
\ee
In eq. (\ref{9}), $\prime$ denotes the derivative with respect to $x$.
The set of all
b.c. can be devided in six (in general overlapping)
different classes, depending on the couple
of quantities, chosen among $\psi(0), \psi(2 \pi), \psi^{\prime}(0)$ and
$\psi^{\prime}(2 \pi)$, with respect to which (\ref{9}) can be solved. We shall
see that in order to get all b.c.
it is in fact sufficient to consider only three cases, which we list
hereafter.\\
{\bf Case 1.} $a_{ij}$ are such that eq. (\ref{9}) can be solved with respect
to $\phi(0)$ and $\phi(2\pi)$:
\be
\pmatrix{\phi^{\prime}(0) \cr \phi^{\prime}(2 \pi)\cr} =
\pmatrix{u & w \cr \zeta & v \cr} \pmatrix{\phi(0) \cr \phi(2 \pi) \cr} ~.
\label{10}
\ee
We show now that
the coefficients $u,v,w$ and $\zeta$ in (\ref{10})
have to fulfill certain conditions.
If (\ref{10}) is to define a s.a. extension of the Laplacian it must be
that if $\psi$ is a $C^2$ function such that:
\be
\langle -\psi^{\prime \prime}, \phi \rangle=
\langle \psi , -\phi^{\prime \prime} \rangle \label{11}
\ee
for all $\phi$ fulfilling (\ref{10}), $\psi$ also has to fulfill (\ref{10}).
Now  we have:
$$
\langle -\psi ^{\prime \prime}, \phi \rangle-
\langle \psi , -\phi^{\prime \prime} \rangle=-\psi^*\phi^{\prime}|_0^{2 \pi}
+ \psi^{\prime *}\phi|_0^{2\pi}=
$$
$$
=-\psi^*(2 \pi)[\zeta \phi(0)+v\phi(2 \pi)]+\psi^*(0)[u \phi(0) +w \phi(2\pi)]
+\psi^{\prime *}(2 \pi)\phi(2\pi)-\psi^{\prime *}(0)\phi(0)=
$$
\be
=[u \psi^*(0)-\zeta\psi^*(2 \pi) - \psi^{\prime *}(0)]\phi(0)+
[w \psi^*(0)-v\psi^*(2 \pi)  + \psi^{\prime *}(0)]\phi(2 \pi)~. \label{12}
\ee
Since now $\phi(0)$ and $\phi(2\pi)$ are arbitrary, $\psi$ must be such
that:
\be
\pmatrix{\psi^{\prime}(0) \cr \psi^{\prime}(2 \pi)\cr} =
\pmatrix{u^* & -\zeta^* \cr -w^* & v^* \cr}
\pmatrix{\psi(0) \cr \psi(2 \pi) \cr}~. \label{13}
\ee
So, by comparing (\ref{12}) with (\ref{10}), we see that $\psi$ fulfills
(\ref{10}) too if and only if:
$$
 u=u^*~~,~~v=v^*~~,~~\zeta=-w^*
$$
and so the most general b.c. of this form can be written as
\be
\pmatrix{\phi^{\prime}(0) \cr \phi^{\prime}(2 \pi)\cr} =
\pmatrix{u & w \cr -w^* & v \cr} \pmatrix{\phi(0) \cr \phi(2 \pi) \cr}~~,~~
u ~~{\rm and} ~~v~~ {\rm real}. \label{14}
\ee
{\bf Case 2.} $a_{ij}$ are such that (\ref{9}) can be solved with respect to
$\phi(0)$ and $\phi^{\prime}(2 \pi)$:
\be
\pmatrix{\phi^{\prime}(0) \cr \phi(2 \pi)\cr} =
\pmatrix{\tilde u & \tilde w \cr \tilde \zeta & \tilde v \cr}
 \pmatrix{\phi(0) \cr \phi^{\prime}
(2 \pi) \cr}. \label{15}
\ee
It is obvious that, if $\tilde v \neq 0$, we can solve (\ref{15}) with respect
to $\phi^{\prime}(2 \pi)$ and the b.c. (\ref{15}) can be rewritten
in the form of case 1. Then, we set $\tilde v=0$. A computation similar to that
performed for case 1, shows that (\ref{15}) define s.a. extensions if and only
if $\tilde w=\tilde \zeta^*$ and $\tilde u$ is real:
\be
\pmatrix{\phi^{\prime}(0) \cr \phi(2 \pi)\cr} =
\pmatrix{\tilde u & \tilde w \cr \tilde w^* & 0 \cr}
\pmatrix{\phi(0) \cr \phi^{\prime}
(2 \pi) \cr}~~,~~\tilde u~~ {\rm real}. \label{16}
\ee
{\bf Case 3.} $a_{ij}$ are such that (\ref{9}) can be solved with respect to
$\phi^{\prime}(0)$ and $\phi^{\prime}(2 \pi)$:
\be
\pmatrix{\phi(0) \cr \phi(2 \pi)\cr} =
\pmatrix{\hat u & \hat w \cr \hat \zeta \ & \hat v \cr}
\pmatrix{\phi^{\prime}(0) \cr \phi^{\prime}
(2 \pi) \cr}. \label{17}
\ee
Again a computation similar to that for case 1 shows that in order to
define a s.a. extension the parameters $\hat u$ and $\hat v$
must be real, while
$\hat \zeta=-\hat w^*$. Then we have:
\be
\pmatrix{\phi(0) \cr \phi(2 \pi)\cr} =
\pmatrix{\hat u & \hat w \cr -\hat w^* & \hat v \cr}
 \pmatrix{\phi^{\prime}(0) \cr \phi^{\prime}
(2 \pi) \cr}~~~\hat u~~{\rm and}~~\hat v~~{\rm real}. \label{18}
\ee
This set of b.c. is not already included in cases 1 and 2
if and only if
$$
\hat u \hat v +|\hat w|^2=0 ~~~{\rm and}~~~\hat u=0.
$$
We are then left with:
\be
\pmatrix{\phi(0) \cr \phi(2 \pi)\cr} =
\pmatrix{0 & 0 \cr 0 & \hat v \cr}
 \pmatrix{\phi^{\prime}(0) \cr \phi^{\prime}
(2 \pi) \cr}~~~~\hat v~~{\rm real}. \label{19}
\ee
It can be checked that the remaining possibilities lead to b.c.
already included in the three cases we have examined. Then, the most
general s.a. extension of the Laplacian on an interval is described
by one of the above sets of b.c.\\

As we have seen earlier, the space of s.a. extensions of the
Laplacian on an interval is a four dimensional manifold, precisely $U(2)$.
Now, for the sake of simplicity and also because later it will allow us
to draw
three-dimensional pictures of the renormalization flow for the discretized
Laplacians, we will restrict our attention to
those extensions which are invariant
under $PT$ symmetry ($P$ here stands for the Parity transformation which
interchanges the end points $x=0$ and $x=2 \pi$ while
$T$ stands for Time-Reversal). Now under $P$ the wave function $\phi$
transforms
as:
\be
(P\phi)(x)=\phi(2 \pi - x) \label{20}
\ee
while under $T$ we have:
\be
(T \phi)(x)= \phi^*(x) \label{21}
\ee
and so under $PT$ we have:
\be
\tilde \phi(x) = (PT \phi)(x)= \phi^*(2 \pi - x)~. \label{22}
\ee
Let us check the constraints implied by the $PT$ symmetry on the b.c.
examined earlier.
{\bf Case 1.} The $PT$ transformed wave function fulfills the b.c.:
\be
\pmatrix{\tilde \phi^{\prime}(0) \cr \tilde \phi^{\prime}(2 \pi)\cr} =
\pmatrix{-v & w \cr -w^* & -u \cr}
 \pmatrix{\tilde \phi(0) \cr \tilde\phi(2 \pi)
 \cr}, \label{23}
\ee
and so $PT$ invariance requires $u=-v$. We then have b.c. of the form:
\be
\pmatrix{\phi^{\prime}(0) \cr  \phi^{\prime}(2 \pi)\cr} =
\pmatrix{u & w \cr -w^* & -u \cr}
 \pmatrix{ \phi(0) \cr \phi(2 \pi)
 \cr}. \label{24}
\ee
{\bf Case 2.} A computation similar to that for case 1, shows that $PT$
invariance requires $|\tilde w|^2=1$. So we have:
\be
\pmatrix{\phi^{\prime}(0) \cr  \phi(2 \pi)\cr} =
\pmatrix{\tilde u & e^{i \theta} \cr e^{-i \theta} & 0 \cr}
 \pmatrix{ \phi(0) \cr \phi^{\prime}(2 \pi)
 \cr}. \label{25}
\ee
{\bf Case 3.} $PT$ invariance requires $\tilde v=0$.
Then we are left just with one set of b.c., those for an open line:
\be
\phi(0)=\phi(2 \pi)=0. \label{26}
\ee

We turn now to the eigenvalue problem for the $PT$ invariant s.a. extensions
listed above. The general
solution to the eigenvalue equation:
\be
-\phi^{\prime \prime}=p^2 \phi \label{27}
\ee
is of the form
\be
\phi_p=A e^{ipx}+B e^{-ipx}. \label{28}
\ee
If $p$ is real the function (\ref{28}) is a plane wave solution
and $p$ represents the inverse wavelength, if $p$ is
imaginary (\ref{28}) represents a bound state solution. An eigenvalue equation
for $p$ is obtained after imposing on (\ref{28}) the b.c. (\ref{24}),
(\ref{25}) or (\ref{26}). We then get the following equations:\\
{\bf Case ~1.} The coefficients $A$ and $B$ have to fulfill the following
linear eqs:
$$
A(u+we^{i 2 \pi p}-ip)+B(u+we^{-i 2 \pi p}+ip)=0
$$
\be
A(u e^{i 2 \pi p}+w^*+ipe^{i 2 \pi p})+B(u e^{-i 2 \pi p}+w^*
-ipe^{-i 2 \pi p})=0~. \label{29}
\ee
The consistency condition for this system is the eigenvalue eq. for $p$:
\be
(p^2 + |w|^2 -u^2)sin 2 \pi p-2u pcos2\pi p-(w+w^*)p=0~~. \label{30}
\ee
{\bf Case ~2.} The coefficients $A$ and $B$ have to fulfill now the
following linear eqs.:
$$
A(\tilde u+ip e^{i\theta} e^{i 2 \pi p}-ip)+B(\tilde u- i p
e^{i\theta} e^{-i 2 \pi p}+ip)=0
$$
\be
A(e^{i 2 \pi p}-e^{-i\theta})+B( e^{-i 2 \pi p}-e^{-i\theta})=0 \label{31}
\ee
and the consistency condition for this system gives the following
eigenvalue equation for $p$:
\be
2pcos 2 \pi p+ \tilde u sin 2\pi p-2p cos\theta=0~. \label{32}
\ee
{\bf Case ~3.} The coefficients $A$ and $B$ have to fulfill the linear
system:
$$
A+B=0
$$
\be
A e^{i 2 \pi p} + Be^{-i 2 \pi p} =0 ~.\label{33}
\ee
The consistency condition now reads:
\be
sin 2 \pi p = 0 ~.\label{34}
\ee

Of particular interest among (\ref{24}), (\ref{25}) and (\ref{26})
are those s.a. extensions characterized by a set
of scale invariant b.c.. As we shall see later they will correspond
to the fixed points of the renormalization flow in the space of
the lattice discretizations of the Laplacian on an interval.
Among the b.c. considered in case 1, the only scale-invariant ones are:
\be
\phi^{\prime}(0)=\phi^{\prime}(2 \pi)=0~. \label{35}
\ee
These are the b.c. for an open line with ``free" ends. Among the b.c.
under case 2, the scale-invariant ones are:
$$
\phi(0)= e^{i \theta} \phi(2 \pi)
$$
\be
\phi^{\prime}(0)= e^{i \theta} \phi^{\prime}(2 \pi)~. \label{36}
\ee
They can be interpreted as b.c. for a circle, threaded by a magnetic flux.
Finally, the only scale-invariant b.c. of type 3 are:
\be
\phi(0)=\phi(2\pi)=0~. \label{37}
\ee
They describe an open line with completely reflecting end points.\\

\sxn{Discretized Laplacians on an interval.}

In the previous Section, we have seen that the Laplacian admits infinite
self-adjoint extensions on the interval $[0, 2 \pi]$, each distinguished
by a different domain ${\cal D}_U$, characterized by a certain set of
b.c. on the wave functions (\ref{9}).

Suppose now we discretize the interval
by dividing it in $N$ subintervals of width $2\pi / N$ and centered
around the points $x_i=(i - \frac{1}{2})\frac{2 \pi}{N},~i=1, \cdots,N$.
Consider now the subspace ${\cal H}^N$ of stepwise functions $\chi(x)$, which
are constant in each of these subintervals:
\be
{\cal H}^N \equiv {\rm span}~\{\chi^N(x)=\chi^N(x_i)~~~{\rm for}~~~
(i-1)\frac{2 \pi}{N}<x<i\frac{2 \pi}{N}~~~i=1, \cdots,N\}. \label{38}
\ee
${\cal H}^N$ is an $N$-dimensional subspace of ${\cal L}^2 ([0, 2\pi])$. It
is clear that, in the limit $N \rightarrow \infty$, ${\cal H}^N$ becomes dense
in ${\cal L}^2 ([0, 2\pi])$. We can represent each function in ${\cal H}^N$
as an $N$-dimensional complex vector: $(\chi^N(i))=(\chi^N(x_i))$.

Consider now the self-adjoint extension of the Laplacian $\Delta_{\rm per}$
acting on periodic wave functions:
\be
\phi(0)=\phi(2 \pi)~~~~~~~\phi^{\prime}(0)=\phi^{\prime}(2 \pi)~ . \label{39}
\ee
(This extension falls unders case 2, when $\tilde u=0$ and $\tilde
w=1$.) As it is well
known, this extension can be approximated with a finite difference
operator acting on the space ${\cal H}^N$:
\be
(\Delta^N \chi^N)(i)=- Z_N^2[\chi^N(i+1)+\chi^N(i-1)-2\chi^N(i)]~~~
i=1, \cdots, N. \label{40}
\ee
Here $N+1 \equiv 1$ and $0 \equiv N$. $Z_N^2$ is an overall
normalization. Let us see how $\Delta_{\rm per}$ is recovered in the
limit $N \rightarrow \infty$. The eigenfunctions of $\Delta_N$ are
easily seen to be equal to:
\be
\chi^N_m(j)=\exp \left\{ im\Bigg(\frac{2 \pi }{N}j-
\frac{\pi}{N}\Bigg) \right\} = \exp (imx_j)~~~~~ m=0, \pm 1,\cdots,\pm
\Bigg[\frac{N-1}{2}\Bigg]~, \label{41}
\ee
where $[a]$ is equal to $a$ if $a$ is integer and to $a-1/2$ if $a$ is
half-integer. The corresponding eigenvalues are:
\be
\lambda_m^N=-2 Z_N^2\Bigg(cos \frac{2 \pi m}{N}-1 \Bigg). \label{42}
\ee
The eigenfunctions are seen to be stepwise approximations of plane
waves. The exponent $m$ can be interpreted as an inverse wavelength
(in units of the interval length). Now the discretized operators
$\Delta^N$ converge to $\Delta_{\rm per}$ in the following sense. Let us
order the eigenstates of $\Delta_{\rm per}$ according to their
eigenvalues $E_i$:
\be
\psi_i~~~~~~E_i \leq E_{i+1}~~~~~i=1, \cdots, \infty. \label{43}
\ee
Let now $r$ an arbitrary integer and consider the operators $\Delta^N$,
with $N \geq r$. Let us order also their eigenfunctions $\chi^N_m(x)$
according to their eigenvalues $\lambda_m^N$:
$$
\chi_m^N(x)~~~\lambda^N_m \leq \lambda^N_{m+1},~~~~m=1,\cdots,N.
$$
The claim is that the first $r$ eigenfunctions $\chi_1^N(x),\cdots
\chi_r^N(x)$ of $\Delta^N$ converge pointwise to the first $r$
eigenfunctions of $\Delta_{\rm per}$, $\psi_1(x),\cdots,\psi_r(x)$.
Moreover it is possible to fix the constants $Z_N$ so that, in the limit
$N \rightarrow \infty$, also the first $r$ eigenvalues $\lambda_m^N$ converge
to the $E_m$. These statements are easily checked in this example.
For instance, if we choose $Z_N=\frac{N}{2 \pi}$, we see that:
\be
\lim_{N \rightarrow \infty} \lambda_m^N=
\lim_{N \rightarrow \infty}-2\Bigg(\frac{N}{2\pi}\Bigg)^2
\Bigg(cos\frac{2 \pi m}{N}-1\Bigg)=
m^2 =E_m. \label{44}
\ee

The question that arises now is if we can recover all self-adjoint
extensions of the Laplacian on the interval as continuum limits of
finite difference operators similar to (\ref{40}).
We have seen in Sec. 1 that the s.a. extensions of the Laplacian
on an interval all have different domains. On the contrary, any finite
difference operator we can write will act on the entire subspace
${\cal H}^N$ for every $N$. It becomes then unclear how the
continuum operators with their different domains can possibly
emerge in the continuum limit.
As we shall see in
the sequel of this paper this is indeed the case. The
operator (\ref{40}) can be generalized in a natural way to give a family of
finite difference operators containing four real parameters. These parameters
will be subject to a renormalization procedure having $N$, the
number of subdivisions of the interval, as control
parameter and this
will give rise to a flow in the
parameter space. The corresponding trajectories will converge
towards certain fixed points. We shall see that each of these
trajectories leads in the limit to a different s.a. extension
of the Laplacian. We shall argue that all s.a. extensions can be
recovered by simply looking at different trajectories.

Let us generalize the finite difference operator (\ref{40}) on an
interval. Notice that, on an interval, formula (\ref{40}) certainly makes sense
for $i=2,\cdots,N-1$, but becomes ill-defined at the endpoints,
$i=1$ and $i=N$. If we then write $\Delta^N$ as a matrix acting
on the vector $\chi^N(i)$ we get:
\be
\Delta^N \chi^N \equiv -Z_N^2
\pmatrix{* & * & * & * & \cdots & * & *\cr
1 & -2 & 1 & 0 &\cdots &\cdots & 0 \cr
0 & 1 & -2 & 1 & 0 &\cdots & 0 \cr
\cdots & \cdots & \cdots & \cdots & \cdots & \cdots & \cdots \cr
0 & \cdots & \cdots & 0 & 1 & -2 & 1 \cr
*& *& \cdots& *& *& *& *& \cr}
 \pmatrix{\chi^N(1) \cr . \cr . \cr . \cr . \cr \chi^N(N)\cr}. \label{45}
\ee
In (\ref{45}) the first and the last raws of the matrix representing
$\Delta^{N}$ are missing. Since there is no natural way of defining $\Delta^N$
at the end points, we will simply take for $\Delta^N$ the most general
hermitian
completion of its corresponding matrix and write:
\be
\Delta^N=-Z_N^2
\pmatrix{a & 1 & 0 & \cdots & \cdots & \cdots & c \cr
1 & -2 & 1 & 0 &\cdots &\cdots & 0 \cr
0 & 1 & -2 & 1 & 0 &\cdots & 0 \cr
\cdots & \cdots & \cdots & \cdots & \cdots & \cdots & \cdots \cr
0 & \cdots & \cdots & 0 & 1 & -2 & 1 \cr
c^*& 0 & \cdots & \cdots & 0 & 1 & b & \cr}
 \pmatrix{\chi^N(1) \cr . \cr . \cr . \cr . \cr \chi^N(N)\cr}. \label{46}
\ee
Here $a$ and $b$ are arbitrary real parameters, while $c$ ia an arbitrary
complex number. Notice that the matrix elements containing the parametrs
only relate the wavefunction $\chi^N$ at the endpoints and seem therefore
related  to boundary conditions. We also remark that $a,b,c$ constitute a set
of
four real paramters, which is also the dimension of the set of $U(2)$
matrices that parametrize the s.a. extensions of the Laplacian.

\sxn{The eigenvalue problem for $\Delta^N$}

In this Section we study the eigenvalue equation for the matrix $\Delta$
(for simplicity of notation we will omit from now on the superscript
$N$ from the symbols for the finite difference operators $\Delta^N$ or the
wavefunctions $\chi^N(x)$):
\begin{eqnarray}
\Delta \chi_{\lambda} &\equiv&
-Z^2\pmatrix{a & 1 & 0 & \cdots & \cdots & \cdots & c \cr
1 & -2 & 1 & 0 &\cdots &\cdots & 0 \cr
0 & 1 & -2 & 1 & 0 &\cdots & 0 \cr
\cdots & \cdots & \cdots & \cdots & \cdots & \cdots & \cdots \cr
0 & \cdots & \cdots & 0 & 1 & -2 & 1 \cr
c^*& 0 & \cdots & \cdots & 0 & 1 & b & \cr}
\pmatrix{\chi_{\lambda}(1) \cr . \cr . \cr . \cr . \cr \chi_{\lambda}(N)\cr}
\nonumber \\
&= &Z^2\lambda
\pmatrix{\chi_{\lambda}(1) \cr . \cr . \cr . \cr . \cr \chi_{\lambda}(N)\cr}.
\label{47}
\end{eqnarray}
The eigenfunctions of $\Delta$ fall into 5 classes, depending on the
eigenvalue $\lambda$. They are listed hereafter:\\
${\bf I)}~$ If $ \lambda < 0 $, we have:
\be
\chi_{\lambda}(n)=A e^{kn}+Be ^{-kn}, \label{48}
\ee
where $\lambda = -2(cosh k -1)~,k>0$.\\
${\bf II)}~$ If  $\lambda = 0 $, we have:
\be
\chi_{0}(n)=A+Bn~. \label{49}
\ee
${\bf III)}~$ If $0 <\lambda < 4 $, we have:
\be
\chi_{\lambda}(n)=A e^{ikn}+Be ^{-ikn}, \label{50}
\ee
where $\lambda=-2(cosk-1)~,~0<k<\pi$.\\
${\bf IV)}~$ If $ \lambda = 4$, we have:
\be
\chi_{4}(n)=(-1)^n(A+Bn)~. \label{51}
\ee
${\bf V)}~$ If $4 < \lambda < \infty$, we have:
\be
\chi_{\lambda}(n)=A \xi^n+B\xi ^{-n}, \label{51a}
\ee
where $\lambda=-(\xi + \xi^{-1} -2)~~,~\xi < -1$.\\

We prove the statement for case V, the proof being similar in the
other cases. Let $\chi_{\lambda}$ be an eigenfunction with eigenvalue
$\lambda > 4$. Let then $\xi$ be the unique real number smaller than -1 such
that $\lambda=-(\xi + \xi^{-1} -2)$ and define $A$ and $B$ to be
such that:
$$
\chi_{\lambda}(1)= A \xi + B \xi^{-1}
$$
\be
\chi_{\lambda}(2)= A \xi^2 + B \xi^{-2}. \label{52}
\ee
Notice that eq.ns (\ref{52}) have a unique solution because
$$
{\rm det}\pmatrix{\xi & \xi^{-1} \cr
                  \xi^2 &\xi^{-2} \cr}=\xi^{-1}-\xi \neq 0~.
$$
By construction, $\chi_{\lambda}$ is of the form (\ref{51a}) for $n=1,2$. We
now
prove by induction that it must be so for all $n$. So, suppose the statement
is true for $n \leq \overline n$, with $\overline n \geq 2$. Then, since
$\chi_{\lambda}$ is an eigenfunction of $\Delta^N$, it must be that:
$$
-[\chi_{\lambda}(\overline n + 1)+\chi_{\lambda}(\overline n - 1)
-2 \chi_{\lambda}(\overline n)]= \lambda \chi_{\lambda}(\overline n)
$$
or
$$
\chi_{\lambda}(\overline n + 1)= (\xi + \xi^{-1}-2)(A\xi^{\overline n}
+B\xi^{-\overline n})-(A\xi^{\overline n-1}
+B\xi^{-(\overline n-1)})+2 (A\xi^{\overline n}
+B\xi^{-\overline n})=
$$
\be
=A\xi^{\overline n+1}+B\xi^{-(\overline n+1)} ~, \label{53}
\ee
which proves the statement.

The physical meaning of the solutions of type I, II and  III
is clear: they represent stepwise approximations of bound states, zero modes
and plane waves respectively. As for the solutions of type IV and
V we shall see in the next section that they ``disappear" from the
spectrum in the continuum limit and are unphysical. Notice also that
the eigenfunctions of type I, III and V can all be
written as:
\be
\chi_{\lambda}(n)=A z^n+ Bz^{-n}~~~~~\lambda \neq 0,4 \label{54}
\ee
where
$$
\lambda=-(z+z^{-1}-2),
$$
$$
z \in {\cal C}\equiv
]-\infty,-1[ ~\bigcup~ \{z=e^{i \theta}~,~
0<\theta<\pi\}~ \bigcup ~]1, \infty[.
$$

Let us derive now the eigenvalue equation for the states of type I, III and
V. Notice first the the wavefunctions (\ref{54}) automatically satisfy the
eigenvalue equation for $n=2,\cdots,N-1$. Then we just have to consider
the endpoints $n=1$ and $n=N$. We find :
$$
A_N[(a+2)z+cz^N -1]+B_N[(a+2)z^{-1}+cz^{-N}-1]=0
$$
\be
A_N[(b+2)z^N+c^*z-z^{N+1}]+B_N[(b+2)z^{-N}+c^*z^{-1}-z^{-(N+1)}]=0. \label{55}
\ee
The condition of compatibility for these eqs. leads to the eigenvalue
equation for $z$:
\be
[\alpha (z^{N-1}-z^{-(N-1)})+\beta (z^{N}-z^{-N})+ \gamma
(z-z^{-1})-(z^{N+1}-z^{-(N+1)})=0 \label{56}
\ee
where we have defined:
$$
\alpha \equiv |c|^2-(a+2)(b+2)
$$
$$
\beta \equiv a+b+4
$$
\be
\gamma \equiv 2({\rm Re}\,c)~. \label{57}
\ee
The real parameters $\alpha, \beta$ and $\gamma$ introduced here are
independent on each other and are only subject to the condition:
\be
4 \alpha + \beta^2 - \gamma^2 \geq 0~. \label{58}
\ee


For the sake of simplicity we will restrict our analysis,
as we did in the continous
case, only to the set of matrices $\Delta$ which are left invariant
by a $PT$ transformation. Now, under $PT$ the vector $\chi(n)$ transforms
as:
\be
(PT\chi)(n)=\chi^*(N-n+1) \label{59}
\ee
and then in order for the matrix $\Delta$ to be $PT$ invariant it must be that
$a=b$:
\be
\Delta \equiv
-Z^2\pmatrix{a & 1 & 0 & \cdots & \cdots & \cdots & c \cr
1 & -2 & 1 & 0 &\cdots &\cdots & 0 \cr
0 & 1 & -2 & 1 & 0 &\cdots & 0 \cr
\cdots & \cdots & \cdots & \cdots & \cdots & \cdots & \cdots \cr
0 & \cdots & \cdots & 0 & 1 & -2 & 1 \cr
c^*& 0 & \cdots & \cdots & 0 & 1 & a & \cr}. \label{60}
\ee
A $PT$ invariant $\Delta$ is thus identified by three real constants and
we can think of it as a point in $R^3$.
In the next Section we will turn to the study of the continuum
limit for the finite difference operators (\ref{60}) and show that in this
limit we recover all $PT$ invariant s.a. extensions of the Laplacian,
eqs. (\ref{24}), (\ref{25}) and (\ref{26}).

\sxn{The continuum limit for $\Delta^N$.}

In this Section we study the continuum limit of the finite difference
operators (\ref{60}).

In the Appendix it is shown that at least $N-2$ eigenstates of $\Delta$
are of the plane-wave type (eq. (\ref{50})). Let us rewrite them:
\be
\chi_k(n)=A e^{ikn}+Be^{-ikn}=
A e^{i\frac{2 \pi p}{N}n}+Be^{-i\frac{2 \pi p}{N}n}
\equiv \chi_p(n),~~~~~~0<k<\pi~. \label{61}
\ee
The number $p\equiv \frac{N k}{2 \pi}$ represents the
inverse wavelength of the state in units of the interval length. In the
Appendix it is shown that the eigenvalues $k$ are distributed essentially
uniformly in the interval $]0,\pi[$, which implies that the first values of
$p$ are of the order of $1,2,\cdots$. The  lowest
eigenstates then have wavelengths of the order of the length of the interval.

Now, our prescription for taking the continuum limit is the following.
Assume we start with a certain set of values for $a$, $c$ and
$N~,(a_0,c_0;N_0)$, or equivalently with a set of values for $\a,~\b$,
$\g$ and $N,~(\a_0,\b_0,\g_0;N_0)$. We then consider the eigenvalue
equation determining the wavelengths of
the ``plane wave" states (\ref{50}), namely eq. (\ref{56}) with
$z=e^{i2 \pi p/N}$:
\be
\a_0 sin 2 \pi p\Bigg(1- \frac{1}{N_0}\Bigg)+\b_0sin 2 \pi p +
\g_0 sin 2 \pi p\frac{1}{N_0}
= sin 2 \pi
p\Bigg(1+\frac{1}{N_0}\Bigg). \label{62}
\ee
Let now:
\be
p^0_1<p^0_2<\cdots<p^0_i~~~~~~~i \le N, \label{62a}
\ee
be its solutions (here we just list the eigenvalues and do not consider the
existance of eventual degeneracies, which, as it is shown in the Appendix
is an  exceptional case).
Our prescription for taking the continuum limit is then  to vary $\a,~\b$ and
$\g$ with $N$ so that the corresponding eigenvalue  equation for the plane wave
states: \be \a(N) sin 2 \pi p\Bigg(1-\frac{1}{N}\Bigg)+\b(N)sin 2 \pi p +
 \g(N) sin 2 \pi p\frac{1}{N}
= sin 2 \pi
p\Bigg(1+\frac{1}{N}\Bigg) \label{63}
\ee
shares with (\ref{62}) as many wavelengths as possible, starting from the
longest one. We shall see that this prescription fixes completely the
dependence of $\a$, $\b$ and $\g$ on $N$. In general, it will be possible
to hold constant only the first three inverse wavelengts, $p^0_1,~p^0_2,~
p^0_3$, but there will be exceptional cases where we will have to consider
the first four of them.

Having determined the evolution of the parameters $\a, \b, \g$ with $N$, we
have to assign now the renormalization condition for
the overall constant $Z_N$ which appears
in front of
$\Delta^N$, eq. (\ref{60}). This constant can be fixed by demanding that
the eigenvalue
$E_1= Z^2_N \lambda_1$
of the first plane-wave state (the one with inverse wavelength $p^0_1$)
takes some
fixed value, independent on $N$. Now, it can be verified, by studying the
eigenvalue equation, that
$p_1$ is always of the order of unity, which implies for $k_1$ a
value of the order of $1/N$. Consequently, we have:
\be
E_1=Z^2_N\lambda_1=-2Z^2_N(cos k_1-1) \approx (Z_N k_1)^2=\Bigg(Z_N \frac{2\pi
p_1}{N}\Bigg)^2 ~.\label{64}
\ee
We see from this formula that $Z_1$ diverges in the continuum limit like
$N$. Notice that the scaling behaviour of $Z_N$ implies
that the eigenvalues of the states of type IV
and V, which are always larger than or equal to $Z_N^2 4$, diverge in the
limit $N \rightarrow \infty$ and disappear from the spectrum. This means
that the states of type $IV$ and $V$ do not have physical meaning.\\

Our renormalization procedure generates a flow in the
space of $PT$ invariant matrices $\Delta^N$, the control
parameter being $N$.  We will then be able to prove
that, depending on the choice of the initial values $\alpha_0$, $\beta_0$,
$\gamma_0$, the discretized Laplacians (\ref{60}) converge in the continuum
limit to any of the $PT$ invariant s.a. extensions  discussed in Sec. 1. This
will be achieved in two steps. First we will  prove that the eigenvalue
equations (\ref{63}) converge in the limit $N \rightarrow \infty$ to
those found for the $PT$ invariant
s.a. of the Laplacian, eqs. (\ref{30}), (\ref{32}) and (\ref{34}).
The wavelength spectra
of the finite difference operators $\Delta^N$ then converge to those of
the s.a. extensions of the Laplacian. Then we prove that also the
eigenfunctions $\chi^N$ converge, in the sense discussed earlier, to those
of the continous $PT$ invariant Laplacians. This will be done by showing
that the linear systems (\ref{55}) which determine $A_N$ and $B_N$
converge to the corresponding systems for the coefficients $A$
and $B$ of the continuum eigenfunctions.

We summarize here the qualitative features of the flow and anticipate what
boundary conditions for the Laplace operator one gets in the continuum limit:
\begin{itemize}
\item We can describe the space of PT-invariant matrices $\Delta^N$ by means
of the three parameters $a$, $b$ , $c$ or, alternatively, by the real numbers
$\alpha$, $\beta$, $\gamma$ defined in (\ref{57}) and subject to condition
(\ref{58}). Hence the parameter space is:
\begin{displaymath}
{\cal P} \equiv \{ (\alpha,\beta,\gamma) \in {\bf R}^3 : 4\alpha + \beta^2 -
\gamma^2 \geq 0 \} ~.
\end{displaymath}
\item The renormalization group has an infinite number of fixed points,
belonging to the sets:
\begin{displaymath}
L_1 \equiv (\alpha=-1,|\beta| \geq 2,\gamma=0)
\end{displaymath}
\begin{displaymath}
L_2 \equiv (\alpha=1, \beta=0, |\gamma| \leq 2) ~.
\end{displaymath}
\item If the initial data $\alpha_0$, $\beta_0$, $\gamma_0$ are chosen to
belong to such lines, in the continuum limit, one recovers the s.a.
extensions of the Laplacian on an interval which correspond to
scale-invariant boundary conditions. In particular the point $P\equiv(-1,2,0)
\in L_1$ corresponds to the s.a. extension (\ref{35}), the other points on
$L_1$ to the s.a. extensions (\ref{37}) and the points on $L_2$ to (\ref{36}).
\item For all initial data not belonging to a certain surface
$S(\alpha_0,\beta_0,\gamma_0,N_0)=0$ in ${\cal P}$, the parameters $\alpha$,
$\beta$, $\gamma$ follow trajectories that all approach the point $P$, for $N
\rightarrow \infty$ (Figure 1). The continuum limit Laplace operators
corresponding to such trajectories are the ones characterized by the non-scale
invariant boundary conditions falling under Case 1.

\item If the initial data are chosen to belong to the surface $S$, the
trajectories approach a point on the line $L_2$ (Figure 2). The corresponding
continuum limit Laplace operators are charcterized by the non-scale invariant
boundary conditions falling under Case 2.
\end{itemize}

Let us show now that our renormalization prescription is well posed and prove
the above statements.

First, consider the case when $\a_0,\b_0,\g_0$ belong to
$L_1$. It is easy  to verify that the eigenvalue equation (\ref{62}) reduces to
\be
sin 2 \pi p =0~~~0<p<\frac{N}{2}~. \label{65}
\ee
The eigenvalues are the integer and half-integer numbers $m$ less than
$N/2$. Consequently the only effect of increasing $N$ is that of
adding shorter wavelengths to the
spectrum, whithout changing the preexisting ones: according to our
prescription these points then represent fixed points of the flow.
Let us look now at the behaviour
of the eigenfunctions $\chi^N_m$. If $\beta_0=2$, we can easily verify,
by expanding
$z=exp(i 2\pi m/N)$ in powers of $1/N$ and
then keeping only the leading terms in eqs. (\ref{55}), that these equations
reduce to those determining the coefficients $A$ and $B$ of the continuum
eigenfunctions for the Laplacian with b.c. given by eq. (\ref{35}). We can
then conclude that the operators $\Delta^N$
associated to the fixed point $P=(-1,2,0)$ converge to the extension
(\ref{35}). If $\b_0\neq 2$, a similar computation shows that in the continuum
limit the  eigenfunctions of the discretized operators converge to those for
the s.a.  extension of the Laplacian with the b.c. given by eqs. (\ref{37}).

Suppose now that the initial point belongs to $L_2$. Eq. (\ref{62}) reduces to:
\be
2cos 2\pi p=\g_0;~~~~~~0<p<\frac{N}{2}~. \label{66}
\ee
Again the only effect of increasing $N$ is just that of adding shorter
wavelengths to the spectrum and then all points of $L_2$ are fixed points.
As to the eigenfunctions $\chi^N_p$, by looking at the limit of eqs.
(\ref{55}), we can check that they
converge to those for the s.a. extension of the  Laplacian having the b.c.
(\ref{36}), with $2 cos \theta = \g_0$.

We consider now the general case. Let $p_1$, $p_2$ and $p_3$ denote the
first three eigenvalues of eq. (\ref{62}) and suppose that the initial values
of
the parameters $\a_0,\b_0,\g_0;N_0$  are such that the entire function of the
complex plane:
\be
h(x;p^0_j)\equiv{\rm det} \parallel sin 2 \pi p^0_j(1-x),~sin 2 \pi p^0_j,
{}~sin 2 \pi p^0_j x \parallel~~~~j=1,2,3 \label{67}
\ee
does not vanish identically in $x$ [$h$ is identically zero if and only if at
least two of the three eigenvalues $p^0_1,p^0_2,p^0_3$ are both integer or
half-integer; this
exceptional case will be examined later]. Since now
$h(x;p^0_j)$ is an entire  function of $x$, its zeros will be isolated points
in the complex plane  and this  implies that there will be at most a finite
number of them in the disc $|x| < 1/N_0$. Now, according to our renormalization
prescription, the running constants $\a(N),\b(N)$ and $\g(N)$ will be chosen to
be the  solutions of the system of three linear equations:
\be
\a(N) sin 2 \pi p^0_j\Bigg(1-\frac{1}{N}\Bigg)+\b(N)sin 2 \pi p^0_j +
 \g(N) sin 2 \pi p^0_j\frac{1}{N}
= sin 2 \pi
p^0_j \Bigg(1+\frac{1}{N}\Bigg). \label{68}
\ee
This system admits a unique solution for all $N>N_0$ [except possibly for
a finite number of values of $N$ such that $1/N$ is a zero of $h(x)$. When
this is
the case we simply neglect these values of $N$]. This proves that our
scheme is well posed.

We will now analize the limiting behaviour of the solutions $\a(N),\b(N)$ and
$\g(N)$. For this purpose, we define two more entire functions of $x$:
\be
f(x;p^0_j)\equiv{\rm det} \parallel sin 2 \pi p^0_j(1-x), ~
sin 2 \pi p^0_j(1+x),
{}~sin 2 \pi p^0_j x \parallel \label{69}
\ee
and
\be
g(x;p^0_j)\equiv{\rm det} \parallel sin 2 \pi p^0_j(1-x),~ sin 2 \pi p^0_j,
{}~sin 2 \pi p^0_j (1+x) \parallel; \label{69a}
\ee
notice that $f(x)$ is an odd function of $x$, while $g(x)$ is even and that
$f,g$ and $h$ all vanish at $x=0$. The
solution of the system (\ref{68}) can now be written, in terms of $f,g$ and $h$
as: \be
\a(x;p^0_j)=-\frac{h(-x;p^0_j)}{h(x;p^0_j)}~, \label{70}
\ee
\be
\b(x;p^0_j)=\frac{f(x;p^0_j)}{h(x;p^0_j)}~, \label{71}
\ee
\be
\g(x;p^0_j)=\frac{g(x;p^0_j)}{h(x;p^0_j)}~. \label{72}
\ee
$\a(x;p^0_j),\b(x;p^0_j)$ and $\g(x;p^0_j)$,
being ratios of entire functions, are thus analytic functions of
the complex plane having at most poles of finite order. We are now
interested in the limits of these functions when $x\rightarrow 0$.
Let us start from $\a(x;p^0_j)$. Let $\tilde n(p^0_j)$
be the order of the first
non vanishing derivative of $h(x;p^0_j)$ for $x=0$ ($\tilde n$
represents also the
order of the zero of $h(x)$ at the origin; a simple computation shows that
$\tilde n(p^0_j)$ is always larger than 1).
Use of the l'Hopital theorem then implies:
\be
\lim_{x \rightarrow 0} \a(x;p^0_j)=-\frac{\lim_{x\rightarrow
0}\frac{d^n}{dx^n}h(-x)}{\lim_{x\rightarrow
0}\frac{d^n}{dx^n}h(x)}=
-(-1)^{\tilde n(p^0_j)}. \label{73}
\ee
The limiting value of $\a$ is then always equal to either 1 or -1.
In order to study the limits of $\b$ and $\g$ let us write the power
expansions of $h,f$ and $g$ around $x=0$:
\be
h(x;p^0_j)= \sum_{n=\tilde n(p^0_j)}^{\infty} h_n(p^0_j)x^n \label{74}
\ee
\be
f(x;p^0_j)= \sum_{n=1}^{\infty} f_{2n}(p^0_j)x^{2n} \label{75}
\ee
\be
g(x;p^0_j)= \sum_{n=0}^{\infty} g_{2n+1}(p^0_j)x^{2n+1}, \label{76}
\ee
where $h_n$, $f_n$, $g_n$ denote the derivative of order $n$ with respect to
$x$ of $h$, $f$, $g$ respecively.

In order to recover, in the continuum limit,
all s.a. extensions of the  Laplacian,
it seems sufficient to consider only two cases $\tilde n (p^0_j)=2,3$.\\
{\bf Case a.} The initial data are such that $\tilde n(p^0_j)=2$. This
is the most general case. $h$ then has a zero of the second order at
the origin. According to (\ref{73}) we then have:
\be
\lim_{x \rightarrow 0} \a(x;p^0_j)=-1~. \label{77}
\ee
It is easy to check now that:
$$
f_2(p^0_j)=2h_2(p^0_j)
$$
and then use of the l'Hopital formula in eq. (\ref{71}) implies:
\be
\lim_{x \rightarrow 0} \b(x;p^0_j)=\frac{f_2(p^0_j)}{h_2(p^0_j)}=2. \label{78}
\ee
As for $\g(x)$ it can be checked that $g_2(p^0_j) \equiv 0$ and then
we find:
\be
\lim_{x \rightarrow 0} \g(x;p^0_j)=\frac{g_2(p^0_j)}{h_2(p^0_j)}=0. \label{79}
\ee
By putting together eqs. (\ref{77}), (\ref{78}) and (\ref{79}), we see that:
\be
{\rm if} ~\tilde n(p^0_j)=2~~~~\lim_{x\rightarrow 0}(\a(x),\b(x),\g(x))=
(-1,2,0)\equiv P. \label{80}
\ee
We saw earlier that the limiting point $P$ is a fixed point of the flow and
that
the discretized operator associated with it converges in the continuum limit
to the s.a. extension of the Laplacian characterized by the b.c. (\ref{35}).
\indent

We show now that the eigenvalue equation (\ref{63}) converges, for
$N\rightarrow
\infty$, to an equation of the form (\ref{30}). For this purpose, let us expand
the r.h.s. of eqs. (\ref{70}), (\ref{71}) and (\ref{72}) in power series of
$x=1/N$:
 \be
\a(x;p^0_j)=-1+H_1(p^0_j)\frac{1}{N}+H_2(p^0_j)\Bigg(\frac{1}{N}\Bigg)^2+\cdots
\label{81}
\ee
\be
\b(x;p^0_j)=2+F_1(p^0_j)\frac{1}{N}+F_2(p^0_j)\Bigg(\frac{1}{N}\Bigg)^2+\cdots
\label{82}
\ee
\be
\g(x;p^0_j)=G_1(p^0_j)\frac{1}{N}+G_2(p^0_j)\Bigg(\frac{1}{N}\Bigg)^2+
\cdots~~~~~. \label{83}
\ee
If we substitue these expansions in (\ref{63}) and keep only the leading terms
in $1/N$ we get:
$$
\Bigg[\Bigg(-2+H_1(p^0_j)\frac{1}{N}+H_2(p^0_j)\frac{1}{N^2}\Bigg)
cos \frac{2\pi p}{N}+ 2 +
F_1(p^0_j)\frac{1}{N}+F_2(p^0_j)\frac{1}{N^2}\Bigg]sin2 \pi p=
$$
\be
=
\Bigg[H_1(p^0_j)\frac{1}{N}cos2 \pi p - G_1(p^0_j)\frac{1}{N}\Bigg]
\frac{2 \pi p}{N}~. \label{84}
\ee
A simple computation shows that $H_1(p^0_j)+F_1(p^0_j)\equiv 0$
and then eq. (\ref{84}) reduces to:
\be
[(2\pi p)^2+H_2(p^0_j)+F_2(p^0_j)] sin 2 \pi p - 2 \pi p [H_1(p^0_j) cos 2
\pi p - G_1(p^0_j)]=0 \label{85}
\ee
This eq. has the same form as (\ref{30}), if we identify:
$$
|w|^2- u^2 = \frac{1}{(2 \pi)^2}(H_2(p^0_j)+F_2(p^0_j))
$$
$$
2u=\frac{1}{2 \pi}H_1(p^0_j)
$$
\be
w+w^*=-\frac{1}{2 \pi} G_1(p^0_j)~~. \label{86}
\ee
Once these identifications are made, it is possible to check, by
susbstituting the expansions (\ref{81}), (\ref{82}) and (\ref{83}) into eqs.
(\ref{55}), that we  recover, to the leading order in $1/N$, eqs. (\ref{29}).
This shows then that the  continuum limit of our discrete operators coincides
with the $PT$ invariant  s.a. extensions of the Laplacian contemplated under
case 1.

By varying $\a_0,\b_0,\g_0$ or equivalently $p^0_j$ we can get essentially
all s.a. extensions of type 1. Consider infact eq. (\ref{30}). Our prescription
for the continuum limit is such that $p^0_j$ remain eigenvalues also in
that limit and then they are roots of eq. (\ref{30}), with $u$ and $w$ given by
eqs. (\ref{86}). Now, conversely,
it can be checked that, if we start from eq. (\ref{30}) and try
determine the coefficients $u$ and $w$ such that $p^0_j$ are solutions, the
answer is in general unique and is given again by eqs. (\ref{86}). So, assuming
that by varying $\a_0,\b_0,\g_0$ we can get all the triples $p^0_j$ that can
be obtained from eq. (\ref{30}) by looking at its first three solutions, we can
conclude that all of s.a. extensions of type 1
can be recovered. There are a only few
exceptional cases that are left: they correspond to the case
when two or more among the $p^0_j$ are integers or half-integers. These last
extensions will be found by analizing the case of $\a_0,\b_0,\g_0$ such
that $h(x;p^0_j)\equiv 0.$\\
{\bf Case b.} Suppose that the imitial data $(\a_0,\b_0,\g_0;N_0)$
are such that $\tilde n (p^0_j)=3$. This is the same as saying that:
\be
h_2(p^0_j)= {\rm det}\parallel -2 \pi p^0_j
cos 2 \pi p^0_j,~sin 2 \pi p^0_j,
{}~ 2 \pi p^0_j  \parallel=0 \label{87}
\ee
and
\be
h_3(p^0_j)= \frac{1}{2}{\rm det}\parallel -(2 \pi p^0_j)^2
sin 2 \pi p^0_j,~sin 2 \pi p^0_j,
{}~ 2 \pi p^0_j  \parallel \neq 0 ~. \label{88}
\ee
These eqs. specify a surface $S=S(\a_0,\b_0,\g_0;N_0)=0$ in the parameter
space.
Now eq. (\ref{87}) implies then that there exist three real numbers $\mu$,
$\nu$
and $\rho$ such that:
\be
\mu 2 \pi p^0_j cos 2 \pi p^0_j +\nu sin 2 \pi p^0_j + \rho 2 \pi p^0_j=0~.
\label{89}
\ee
Moreover eq. (\ref{88}) implies that $\mu \neq 0$.
$h(x)$ now has a zero of the third order at $x=0$ and eq. (\ref{73}) implies
that:
\be
\lim_{x \rightarrow 0} \a(x;p^0_j)=1~. \label{90}
\ee
A simple computation also shows that:
\be
\lim_{x \rightarrow 0} \b(x;p^0_j)=0 \label{91}
\ee
\be
\lim_{x \rightarrow 0} \g(x;p^0_j)=-2\frac{\rho}{\mu}~. \label{92}
\ee
Condition (\ref{58}) then implies that $|\rho / \mu| \leq 1$.

This computation
shows that the trajectories starting from a point of $S$ will approach the
point $(\a,\b,\g)=(1,0,-2 \rho/\mu)$ belonging to the set $L_2$.
We showed earlier that the points of this set represent fixed points and
that in the continuum limit they correspond to the s.a. extensions of type
2, characterized by the b.c. (\ref{36}). If we now
expand the r.h.s. of eqs. (\ref{70}), (\ref{71}), (\ref{72}) around $x=0$ we
find: \be
\a(x;p^0_j)=1+\tilde H_1(p^0_j)\frac{1}{N}+
\tilde H_2(p^0_j)\Bigg(\frac{1}{N}\Bigg)^2+\cdots \label{93}
\ee
\be
\b(x;p^0_j)=\tilde F_1(p^0_j)\frac{1}{N}+\tilde
F_2(p^0_j)\Bigg(\frac{1}{N}\Bigg)^2+\cdots \label{94}
\ee
\be
\g(x;p^0_j)=-2 \frac{\rho}{\mu}+\tilde
G_1(p^0_j)\Bigg(\frac{1}{N}\Bigg)+
\cdots~~~~~. \label{95}
\ee
Notice that these expansions do not coincide with those conputed earlier,
under case 1. If we substitute these expansions in the eigenvalue eq.
(\ref{63}) and keep only the leading order terms in $1/N$ we get the limiting
eigenvalue eq.:
\be
[\tilde H_1(p^0_j)+\tilde F_1(p^0_j)] sin 2 \pi p - 4 \pi p \Bigg[ cos 2
\pi p + \frac{\rho}{\mu}\Bigg]=0~. \label{96}
\ee
As a consequence of eq. (\ref{89}), we have the following relations:
$$
\tilde H_1(p^0_j)=-\frac {2\nu}{3\mu}
$$
$$
\tilde F_1(p^0_j)=-\frac {\nu}{3\mu}
$$
\be
\tilde G_1(p^0_j)=\frac {2 \rho \nu}{3\mu^2} \label{97}
\ee
and then eq. (\ref{96}) can be rewritten as:
\be
\frac{\nu}{\mu} sin 2 \pi p + 4 \pi p \Bigg[cos 2
\pi p + \frac{\rho}{\mu}\Bigg]=0~~~~~~\mu \neq 0 \label{98}
\ee
which is of the same form as eq. (\ref{32}), if we identify:
\be
\tilde u = \frac{\nu}{ 2 \pi \mu} \label{99}
\ee
\be
cos \theta=-\frac{\rho}{\mu}~. \label{100}
\ee
Using these identifications, we will now prove that,
in the continuum limit, the linear system (\ref{55}) is
equivalent to its analogue, eqs. (\ref{31}), for the $PT$ invariant extensions
contemplated under case 2. Notice first that, to the zeroth order in
$1/N$, eqs. (\ref{55}) are both equivalent to the following eq.:
\be
A(e^{i \theta} e^{i 2 \pi p}-1) + B (e^{i \theta} e^{-i 2 \pi p}-1)=0
\label{101}
\ee
which is clearly equivalent to the second of eqs. (\ref{31}). If we now
subtract
the second of eqs. (\ref{55}) from the first and expand the difference to the
first order in $1/N$ and use eqs. (\ref{97}), we find:
\be
A\Bigg(\frac{\nu}{2 \pi \mu}+ ip e^{-i \frac{\rho}{\mu}}e^{i 2 \pi p}-ip
\Bigg)+
B\Bigg(\frac{\nu}{2 \pi \mu} - ip e^{-i \frac{\rho}{\mu}}e^{-i 2 \pi p} + ip
\Bigg)=0 \label{102}
\ee
which, in view of eqs. (\ref{99}) and (\ref{100}), is equivalent to the
first of eqs. (\ref{31}).\\

We now turn to the exceptional case, when the initial data $(\a_0, \b_0,
\g_0;N_0)$ are such that $h(x;p^0_j) \equiv 0$. This case occurs if and only
if at least two of the eigenvalues $p^0_j$ are both integer or
half-integer and this happens if and only
if $\a+1+\g=0$. In this case, $p^0_1$ and $p^0_3$ are actually half-integer,
while $p^0_2$ is neither integer nor half-integer.
The plane of equation $\a+1+\g=0$ is then invariant under the
renormalization flow , but in order to completely fix the
evolution of the running constants $\a(N)$, $\b(N)$ and $\g(N)$ we have to
consider also the fourth eigenvalue $p^0_4$, which again is neither integer
nor half-integer. If we use the relation $\a+1+\g=0$ to eliminate $\g$, the
evolution of $\a$ and $\b$ is then determined by the equations:
\be
\a(N)[ sin 2 \pi p^0_l (1-x) - sin 2 \pi p^0_l x]+ \b(N)sin 2 \pi p^0_l=
sin 2 \pi p^0_l (1+x) + sin 2 \pi p^0_l x ~~~~l=2,4 \label{102a}
\ee
(in what follows we will not show again the range of values for the index
$l$; it will be understood that $l=2,4$). We can apply  to these eqs. the
same methods as those used for eqs. (\ref{63}), and the conclusions are
similar.  Therefore we will be brief here. We introduce the entire functions:
\be
s(x;p^0_l)\equiv det \parallel
sin 2 \pi p^0_l (1-x) - sin 2 \pi p^0_l x,~ sin 2 \pi p^0_l \parallel
\label{103}
\ee
\be
t(x;p^0_l) \equiv det \parallel
sin 2 \pi p^0_l (1-x) - sin 2 \pi p^0_l x,~sin 2 \pi p^0_l (1+x) +
 sin 2 \pi p^0_l x \parallel. \label{104}
\ee
Notice that both these functions vanish for $x=0$ and that $t(x;p^0_l)$ is an
odd function of $x$. Using  these functions we can write the solution to eqs.
(\ref{102a}) as: \be
\a(x;p^0_l)=\frac{s(-x;p^0_l)}{s(x;p^0_l)}~, \label{105}
\ee
\be
\b(x;p^0_l)=\frac{t(x;p^0_l)}{s(x;p^0_l)}~. \label{106}
\ee
Let now $\tilde r(p^0_l)$ be the order of the first non-vanishing derivative
of $s(x)$ for $x=0$. Use of l'Hopital theorem implies:
\be
\lim_{x \rightarrow 0} \a(x;p^0_l)=(-1)^{\tilde r(p^0_l)}. \label{107}
\ee
As before, it is sufficient to consider two cases.\\
{\bf Case a:} $\tilde r(p^0_l)=1$. A simple computation analogous to that
done earlier shows that:
\be
\lim_{x \rightarrow 0}(\a(x;p^0_l),\b(x;p^0_l),\g(x;p^0_l))=(-1,2,0)~.
\label{108}
\ee
Moreover, the eigenvalue equation, eq. (\ref{56}) and the linear system
determining $A$ and $B$ , eqs. (\ref{55}), reduce in the continuum limit to the
corresponding equations for the Laplacian with b.c. of type 1.\\
{\bf Case b:} $\tilde r(p^0_l)=2$. This case occurs if and only if:
$$
s_1(p^0_l)=\frac{d s(x;p^0_l)}{dx}\Bigg|_{x=0}=
 - det \parallel 2 \pi p^0_l(cos 2 \pi p^0_l+1),~sin 2 \pi p^0_l
\parallel=
$$
\be
-4 cos  \pi p^0_2 cos  \pi p^0_4
det \parallel 2 \pi p^0_l cos  \pi p^0_l,~sin \pi p^0_l
\parallel=0~. \label{109}
\ee
This conditions is fulfilled if and only if there exist real numbers
$\s$ and $\tau$ such that:
\be
2 \pi \s p^0_l sin \pi p^0_l + \tau cos \pi p^0_l=0~. \label{110}
\ee
Computations
similar to the previous ones show that:
\be
\lim_{x \rightarrow 0}(\a(x;p^0_l),\b(x;p^0_l),\g(x;p^0_l))=(1,0,-2)~.
\label{111}
\ee
The limits of the eigenvalue eq. (\ref{56}) and of the linear system (\ref{55})
coincide now with those for the Laplacian with b.c. of type 2.
\newpage

\centerline {{\bf APPENDIX}}

\axn{Study of the eigenvalue equation for $\Delta^N$}

In this Appendix we analize the eigenvalue equation for the finite
difference operators $\Delta^N$. We start by showing that at least $N-2$
eigenfunctions are of type III, (eq. (\ref{50})), namely of the plane-wave
type. Consider the eigenvalue eq. (\ref{56}), for $z=e^{ik}$ with $0<k<\pi$:
\be
\a sin k(N-1) + \b sin k N+ \g sin k - sink(N+1)=0~. \label{112}
\ee
$$
0<k<\pi~.
$$
It is convenient to expand the sines in eq. (\ref{112}) and rewrite it as:
\be
[(\a -1)cos k +\b]sin kN=[(\a+1)cos kN-\g]sin k~. \label{113}
\ee
There are now several cases. Here we will examine only some of them, the
analysis being similar in the other cases. First, there are some exceptional
cases.

{\bf Case a:} $\a=-1$ and $\g=0.$ Condition (\ref{58}) then implies $|\b| \geq
2$.  These points all belong to the line $L_1$. In this exceptional case eq.
(\ref{112}) simplifies to
\be
sinNk=sin 2 \pi p=0 ~.\label{114}
\ee
The eigenvalues $p$ are the integers and the half-integers less then $N/2$
and there are
$N-1$ of them. It can be checked that $N/2$ is also an eigenvalue.

{\bf Case b:} $\a=1$ and $\b=0$. Condition (\ref{58}) implies $|\g| \leq 2$.
These points all belong to the line $L_2$. In this other exceptional case
eq. (\ref{112}) reduces to:
\be
2coskN= cos 2 \pi p=\g~. \label{115}
\ee

{\bf Case c:} $\a+1 \pm \g = 0~;~\a-1 + \b\neq 0$ and $\b$ and $\g$ are not
such that we are in cases  (a) and ( b). Now $kN= (2m
-1)\pi,m=1,\cdots[N/2]$ are solutions to (\ref{113}). The remaining solutions
are  the roots of the equation:
\be
\frac{(\a-1)cos k +\b}{sin k}=(\a+1)cotg\frac{kN}{2}~. \label{116}
\ee
This equation always has exactly one solution in each interval $\frac
{\pi}{N}(2 m-1)<k<\frac {\pi}{N}(2 m+1), m=1, \cdots,[N/2]$ for $N$ odd and
$m=1, \cdots,N/2-1$ for $N$ even. These solutions are neither integer nor
half-integer. All in all we have then $N-1$ plane wave solutions.

{\bf Case d:} $\a+1 \pm \g \neq 0$ (and the values of $\a, \b$ and
$\g$ are such
that we are not in case (b). This is the most general case.
Now $p$ cannot be integer or
half-integer and then eq. (\ref{112}) is equivalent to:
\be
\frac{(\a-1)cosk~+~\b}{sin k}=\frac{(\a+1)cos kN~-~\g}{sin kN}~. \label{117}
\ee
There are now several subcases, depending on the relative signs of
$\a-1 \pm \b$ and $\a+1 \pm \g$. We will discuss here only two subcases, the
analysis being similar in the others. So suppose that:
\be
\b > |\a-1|~~~~~~~~~~|\g|<\a+1 ~.\label{118}
\ee
Condition (\ref{58}) is then always satisfied. The l.h.s. of eq. (\ref{113}) is
then always positive for $0<k<\pi$ and diverges both at $k=0$ and $k=\pi$.
The r.h.s. is a periodic function with period $2 \pi/N$. Its qualitative
behaviour is the same as that of the function $cotg kN$ and it is easy to
verify that there is exactly one solution in each of the intervals
$\frac{\pi}{N}n<k<\frac{\pi}{N}(n+1),~n=1,\cdots, N-1$. There are then at
least $N-1$ plane wave solutions.

It is easily verified that there is also one
solution of type I, namely a bound state. Since we have managed to find $N$
eigenvalues, it also follows that they are all non degenerate.

Consider now the case:
\be
\b > |\a-1|~~~~~~~~~~\g>|\a+1|~, \label{119}
\ee
where $\a,~\b$ and $\g$ have to fulfill condition (\ref{58}). The l.h.s.
has  the same behaviour as in the previous subcase, but the r.h.s. has the
qualitative behaviour of the function $-\g / sin kN$. It can be checked that
the conditions (\ref{58}) ensures that the minimum value of the l.h.s. is
larger  than the minimum value of the r.h.s.. It is then easy to verify that
there are exactly two distinct eigenvalues in each of the subintervals
$\frac{\pi}{N}(2m-1)<k<\frac{\pi}{N}2m,~m=1,\cdots [N/2]-1$. Moreover, if
$N$ is even there
is one more solution in the interval $\frac{\pi}{N}(N-1)<k<\frac{\pi}{N}$.
For all $N$ there are then $N-1$ solutions.

Again, it can be checked that there exists one solution of type I. As
before, these eigenvalues are then all non-degenerate.\\

We show now that eq. (\ref{113})
admits degenerate solutions only if $\a, \b, \g$ belong
to a a certain curve. From eqs. (\ref{55}),with $a=b$ and $z=e^{ik}$,
we see that $k$ is a degenerate
eigenvalue if and only if the coefficients of $A_N$ and $B_N$ all vanish
simultaneously. This condition is equivalent to:
$$
(a+2)e^{ik}+ce^{ink}=1
$$
\be
(a+2)e^{-ik}+ce^{-ink}=1~. \label{120}
\ee
This system is compatible if and only if $sin(N-1)k\neq 0$. Then its
solution is:
$$
a+2=\frac{\b}{2}=\frac{sin Nk}{sink(N-1)}
$$
\be
c=- \frac{sink}{sink(N-1)}~. \label{121}
\ee
When $k$ is varied between $0$ and $\pi$
in eqs. (\ref{121}) $\a,\b,\g$ describe a curve: this is then the set of values
of the parameters for which some of the eigenvalues of eq. (\ref{112}) can be
degenerate.

\newpage
\begin{center}
\begin{large}
{\bf Aknowledgements}
\end{large}
\end{center}

We thank A.P. Balachandran for encouraging us to look into this problem and
providing many useful suggestions. We thank also L. Chandar for several
fruitful discussions.

This work was supported in part by the Department of Energy, U.S.A., under
contract number DE-FG02-ER40231.

\end{document}